\documentclass[english,aps,pra,superscriptaddress,twocolumn]{revtex4-1}
\usepackage[T1]{fontenc}
\usepackage{natbib}
\usepackage[latin9]{inputenc}
\usepackage{graphicx}
\usepackage{babel,bm}
\usepackage{amsfonts}
\usepackage{amsmath}
\usepackage{amssymb}
\usepackage{graphicx}
\usepackage{color}
\usepackage{epstopdf}
\usepackage{hyperref}
\usepackage{txfonts}
\usepackage{enumerate}

\begin{document}

\title{Practical security analysis of two-way quantum key distribution protocols
based on non-orthogonal states }

\author{C. Ivan Henao and Roberto M. Serra}

\affiliation{Centro de Ci\^{e}ncias Naturais e Humanas, Universidade Federal do ABC,
Av. dos Estados 5001, 09210-580 Santo Andr\'{e}, S\~{a}o Paulo, Brazil}

\begin{abstract}
Within the broad research scenario of quantum secure communication, Two-Way
Quantum Key Distribution (TWQKD) is a relatively new proposal for
sharing secret keys that is not fully explored yet. We analyse the security of 
TWQKD schemes that use qubits prepared in non-orthogonal states to 
transmit the key. Investigating protocols that employ
an arbitrary number of bases for the channel preparation, we show, in particular,
that the security of the LM05 protocol can not be improved
by the use of more than two preparation bases. We also provide
a new proof of unconditional security for a deterministic
TWQKD protocol recently  proposed [Phys. Rev. A \textbf{88},
062302 (2013)]. In addition, we introduce a novel deterministic protocol named ``TWQKD six-state''  
and compute an analytical lower bound (which can be tightened) for the maximum amount of information 
that an eavesdropper could extract in this case. An interesting advantage of our approach to the security analysis of TWQKD 
is the great simplicity and transparency of the derivations. 
 
\end{abstract}
\maketitle

\section{{INTRODUCTION}}

Quantum Key Distribution (QKD) harnesses the laws of Quantum Mechanics
to distribute a secret key with a security level unachievable by classical
means \cite{gisin02, scarani09,koashi14,vidick14}. In a QKD protocol two parties, commonly called Alice 
(message sender) and Bob (message receiver), want to establish a secret key between them, by sending 
quantum and classical information through an insecure channel. The communication channel can be spied 
by a powerful eavesdropper, typically called Eve, who is assumed to be technologically much more 
advanced than Alice and Bob. Thus, Eve can listen all the transmitted classical messages and manipulate 
the quantum information at her own will, in principle, being only limited by the laws of Quantum Physics. 
However, this manipulation unavoidably introduces perturbations of quantum nature, which may be detected by the communication partners
(Alice and Bob), depending on the features of the communication protocol they are employing. In such a way, they are able to determine how much information was leaked and keep only the secure part in the final secret key.
 
Standard QKD protocols (One-Way protocols), such as BB84 \cite{key-1,key-2-1,key-3-1,key-19,key-3-1.1}, six-state \cite{key-2,key-34}
and and SARG04 \cite{key-19,key-16.1,key-16,key-18}, use an encoding method that prevents Bob to decode the information in a deterministic way. This means that a fraction of the 
transmitted key bits must be discarded, due to the fact that Bob's measurements do not allow him to deduce with certainty the corresponding values. 
On the other hand, Two-Way Quantum Key Distribution (TWQKD) protocols \cite{key-3,key-4,ali11,shaari12,key-5,key-5.1,key-5.2} can provide a solution to this flaw. 
In this kind of protocols the preparation of the quantum states that will be encoded is carried out by Bob. Therefore, with this additional knowledge, he can always perform the 
right decoding measurements and no bits must be discarded. An important factor to assess the performance of a given QKD protocol is the key
rate, that is, the number of secret key bits distributed per unit
of time. Besides the improvements related to technological developments
\cite{key-6}, the use of qudits instead of two-level systems 
has been proposed as an option to achieve larger secret key rates \cite{key-7}.
TWQKD protocols have also the potential to fulfill this goal, given
that the key bits can be decoded more efficiently than with One-Way
schemes. 

Regarding security, it was recently shown that the performance of some TWQKD protocols is comparable to that of their One-Way counterparts 
and can even surpass it \cite{key-5}. The two paradigmatic TWQKD schemes proposed
to date are LM05 (based on non-orthogonal states) \cite{key-4} and
the Ping-Pong protocol (based on entanglement) \cite{key-3}, both
of them being deterministic. In spite that the original Ping-Pong
protocol is vulnerable to zero error attacks \cite{key-8,key-9},
there exists a non-deterministic version that overcomes this shortcoming
and has been proven secure \cite{key-5}. On the other hand, security
proofs for LM05 and a deterministic TWQKD protocol inspired on it
(LM05') are given in \cite{key-10,key-11} and \cite{key-5}, respectively. 

In this work we investigate the security of several TWQKD schemes
based on non-orthogonal states, including deterministic and non-deterministic
ones. The proofs that we construct guarantee security against collective
attacks, when the classical post-processing is implemented with ``direct
reconcilliation'' \cite{scarani09}. In Section II
we describe the basic structure of the protocols studied here. In
Section III we present a simple TWQKD protocol, whose security proof
constitutes the basis of some subsequent results. Next, we prove in
Section IV the security of the LM05' protocol, by introducing a slight
modification that reduces the corresponding analysis to that of the
scheme described in Section III. Finally, Section V is focused on
the study of two new deterministic TWQKD protocols. One of them can
be seen as a generalization of the LM05 protocol, due to the use of
more than two preparation bases. We prove that the security of this generalized
protocol is the same of LM05. For the latter protocol, which we term
``TWQKD six-state'', we provide a lower bound for the eavesdropping
information that coincides with the corresponding to the six-state
protocol \cite{scarani09,key-15,key-17}. The security of the novel TWQKD six-state protocol is at least as good as that of LM05' and probably even bigger (considering the tightness of the bound tracked down). All the calculations are performed
analytically, showing explicitly the attacks that allow Eve to get
a given amount of information. We take advantage of the so called Gram matrix formulation \cite{key-27} in our analysis.

\section{General structure of the Two-Way protocols investigated}\label{section:TWQKD}

\begin{figure}[t]
\includegraphics[scale=0.0405]{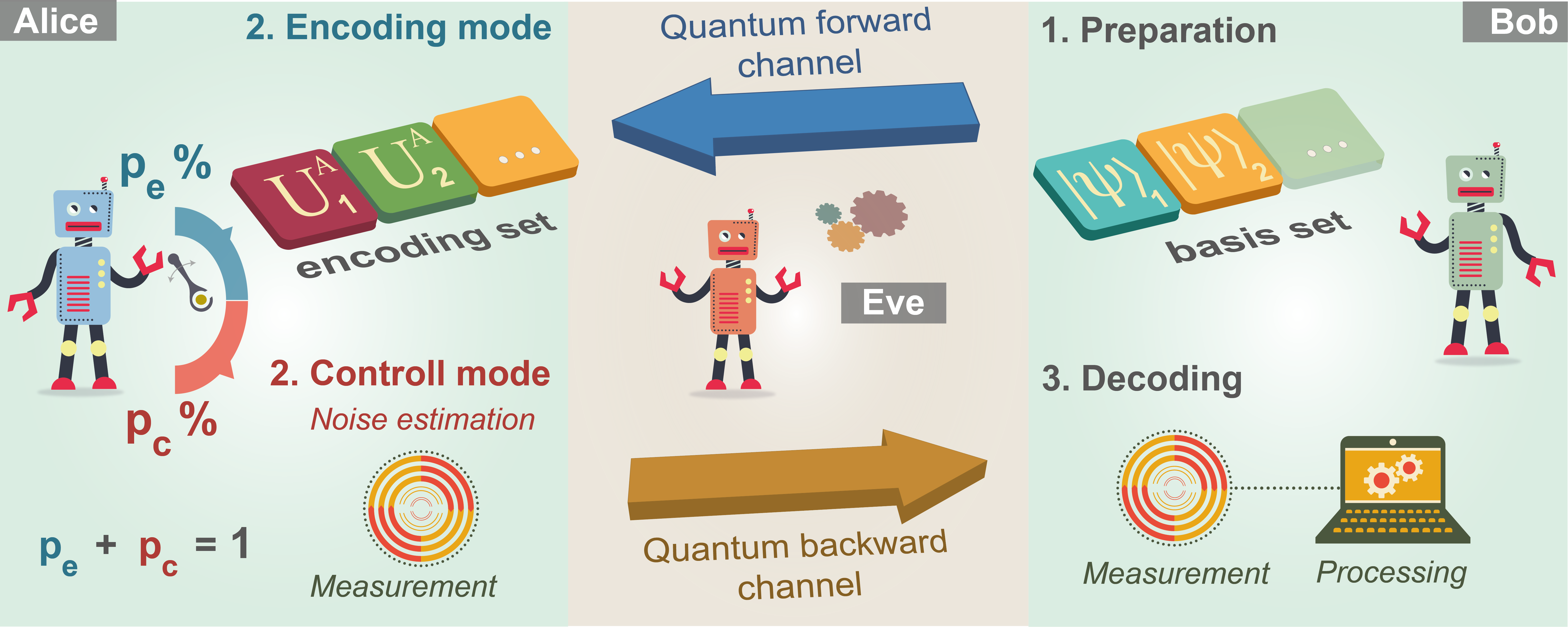}
\caption{(Color online) Schematic illustration of the quantum part of the general TWQKD
protocol, with Alice at the left side, Bob at the right side and a eventual eavesdropper (Eve) at the center (the classical communication channel is omitted). The
states $\{|\psi_{i}\rangle\}_{i}$ and the unitary operations $\{U_{i}^{A}\}_{i}$
are chosen probabilistically by Bob and Alice, respectively. Bob performs the decoding measuring the qubit
in the preparation basis and for CM Alice measures the received
qubit in any of the possible preparation bases (randomly chosen). Alice also chose randomly  between EM and CM with probabilities $p_{e}$ and $p_{c}$, respectively. See the general protocol description in the main text for more details.}
\label{fig:1}
\end{figure}

The basic modus operandi of the protocols considered here is the following
(see Fig.~\ref{fig:1} for a general description): 
\begin{enumerate}
\item Bob (the message receiver) prepares the $i$-th qubit state from a basis set $\{|\psi_{i}\rangle\}_{i}$,
according to some probability distribution, and sends it to Alice
through the \textit{quantum forward channel} (QFC).
\item Alice (the message sender) performs the Encoding Mode (EM) with probability $p_{e}$ and Control
Mode (CM) with probability $p_{c}$ ($p_{e}+p_{c}=1$). In EM she
encodes one bit of information, by randomly applying unitary operations
from a predetermined set $\{U_{i}^{A}\}_{i}$, and sends back the
qubit to Bob through the \textit{quantum backward
channel} (QBC). Instead, in CM, she measures the received qubit in a given basis, by
choosing randomly among the possible preparation bases used by Bob,
to get information about the prepared state. It is important to note that 
the QFC and the QBC are the same physical channel
and the distinction is used to facilitate the description of the eavesdropping
attacks. 
\item Bob measures the qubit in the $i$-th preparation basis to decode the information. 
\item After all the qubits have been transmitted and measured,
a leftover part of the protocol is carried out throughout a classical channel.
Alice publicly reveals which qubits were used for CM and which ones
for EM. In the case of CM Alice and Bob compare the prepared states
and the measurement outcomes, whenever Alice's measurements had been performed
properly in the preparation basis. This procedure allows them to establish
the \textit{forward noise} $Q_{f}$, defined as
the probability that such states don't match up. On the other hand,
a fraction of the systems used for EM is randomly chosen to determine
the probability that encoded and decoded symbols be different, $Q$, the overall noise.
The remaining symbols are not disclosed and constitute what is known
as the ``raw key''.
\item \textit{Classical post-processing}: Alice and Bob
interchange additional classical information in order to obtain identical
and completely secret bit strings starting from their raw keys. To
this aim two sub-protocols are required. The first is called ``error
correction'' \cite{key-12} and it is used to amend the errors associated
to the overall noise $Q$. In this way, the encoded and decoded keys become
perfectly correlated, i.e., Alice's and Bob's bits are identical. In
the following sub-protocol information potentially leaked during the
quantum transmission is removed from the key, which is known as ``privacy
amplification'' \cite{key-13}. Here we will consider the usual approach
for the classical post-processing, where the involved sub-protocols
are implemented with one-way classical communication (see, e.g., \cite{key-14}
for a different approach). 
\end{enumerate}

\section{Security analysis for a simple TWQKD protocol}

\subsection{General eavesdropper attack and security proof}

Let us consider a generic TWQKD protocol where Alice uses the encoding
operations set $\{\mathbb{I}^{A},\sigma_{z}^{A}\}$ (the bits ``0'' and ``1''
are encoded with the application of the unitary operation $\mathbb{I}^{A}$ and $\sigma_{z}^{A}$, respectively)
and Bob's preparation is such that Eve observes the state $\mathbb{I}^{A}/2$
in the QFC. In what follows we will denote operators acting on the
encoded qubit Hilbert space with the superscript $A$. Other details, such
as the preparation bases used by Bob, are irrelevant by now. For keys
of infinite size, the secret fraction, $r$, is given by \cite{key-21,key-22}
\begin{equation}
r=I(A:B)-I_{E}=1-h(Q)-I_{E},\label{eq:1}
\end{equation}
where $I(A:B)$ is the mutual information between Alice's and Bob's
data, which measures the correlation of the encoded and decoded keys.
If Alice sends one bit, i.e., she uses $\mathbb{I}^{A}$ and $\sigma_{z}^{A}$
with equal probability, $I(A:B)=1-h(Q)$, being $h(x)=-x\log_2(x)-(1-x)\log_2(1-x)$ the binary
entropy and $Q$ the probability of miss-matching between the encoded and decoded bit. 
Here, we assume (as usual) that the quantum channel is a depolarizing channel
\cite{key-23} and hence $Q$ does not depend on
the specific states carrying the information. Moreover, $I_{E}$ represents
the maximum information that the eavesdropper (Eve) can obtain using
her best strategy (the expression for $I_{E}$ will be presented latter).

We consider the case of collective attacks and classical post-processing
implemented with ``direct reconciliation'', implying that $I_{E}$
corresponds to Eve's knowledge about the encoded key \cite{scarani09}.
Accordingly, the most general attack that Eve can perform in the QFC
consists of a joint interaction between the sent qubit and some auxiliar
ancilla. Without loss of generality, this interaction may be represented
as 
\begin{equation}
U|i\rangle_{\hat{k}}|\epsilon\rangle=|0\rangle_{\hat{k}}|\epsilon_{i0}^{\hat{k}}\rangle+|1\rangle_{\hat{k}}|\epsilon_{i1}^{\hat{k}}\rangle;\quad i=0,1.\label{eq:2}
\end{equation}
where $U$ is a unitary operation acting on the joint Hilbert space
of the ancilla and the qubit, and $|\epsilon\rangle$ is the ancilla
state before the attack. The set $\{|i\rangle_{\hat{k}}\}_{i=0,1}$
contains the eigenstates of the Pauli operator $\hat{k}.\vec{\sigma}$,
which corresponds to the $\hat{k}$ direction in the Bloch sphere.
Unitarity is guaranteed whenever the ancilla states $\{|\epsilon_{ij}^{\hat{k}}\rangle\}_{0\leq i,j\leq1}$
fulfil the conditions: 
\begin{equation}
\langle\epsilon_{i0}^{\hat{k}}|\epsilon_{i0}^{\hat{k}}\rangle+\langle\epsilon_{i1}^{\hat{k}}|\epsilon_{i1}^{\hat{k}}\rangle=1,\quad\langle\epsilon_{00}^{\hat{k}}|\epsilon_{10}^{\hat{k}}\rangle+\langle\epsilon_{01}^{\hat{k}}|\epsilon_{11}^{\hat{k}}\rangle=0.\label{eq:2.1}
\end{equation}

After the forward attack, Eve resends the qubit to Alice and stores
her ancilla until the qubit is sent back through the QBC. In this
way, the joint encoded states that she can access in the QBC are $\rho^{AE|0}=U(\mathbb{I}^{A}/2\otimes|\epsilon\rangle\langle\epsilon|)U^{\dagger}$
and $\rho^{AE|1}=\sigma_{z}^{A}U(\mathbb{I}^{A}/2\otimes|\epsilon\rangle\langle\epsilon|)U^{\dagger}\sigma_{z}^{A}$, with $\rho^{AE|0}$ ($\rho^{AE|1})$ being the state conditioned to the Alice's encoded bit ``0'' (``1''). Following the definition of collective attack, Eve can extract the
encoded information by performing coherent measurements on any number
of qubits and ancillae that she considers convenient. To this aim,
it is assumed that she has an unlimited resource of quantum memory
at her disposition \cite{key-25}. Accordingly, the expression for Eve's
information is given by \cite{key-22}: 
\begin{align}
I_{E} & =\underset{\{U\}}{\mathrm{max}}\, \chi \nonumber \\
& =\underset{\{U\}}{\mathrm{max}}\left\{S\left(\frac{\rho^{AE|0}+\rho^{AE|1}}{2}\right)-\frac{1}{2}[S(\rho^{AE|0})+S(\rho^{AE|1})]\right\},\label{eq:3}
\end{align}
where $S(\rho)=\operatorname{Tr} \rho \log_2 \rho$ is the von Neumann entropy and $\chi$ is the Holevo quantity
\cite{key-26} for the alphabet encoded into $\{\rho^{AE|i}\}_{i=0,1}$. The maximization is taken over all possible Eve's unitaries.

Since the conditional density operator $\rho^{AE|i}$ only differ of $\mathbb{I}^{A}/2\otimes|\epsilon\rangle\langle\epsilon|$
by unitary transformations, we can write $S(\rho^{AE|i})=S(\mathbb{I}^{A}/2\otimes|\epsilon\rangle\langle\epsilon|)=1$.
Hence, 
\begin{equation}
I_{E}=\underset{\{U\}}{\mathrm{max}} \, \left[ S(\rho^{AE})-1 \right],\label{eq:4}
\end{equation}
with $\rho^{AE}=\frac{1}{2}(\rho^{AE|0}+\rho^{AE|1})$.

To compute Eve's information, in Eq.~\eqref{eq:4}, we need first to obtain the eigenvalues of the Alice-Eve joint state $\rho^{AE}$. We will perform this task employing the Gram matrix representation \cite{key-27}. For a mixed state (written as a mixture of non-orthogonal pure states) $\rho=\sum_{i}p_{i}|\varphi_{i}\rangle\langle\varphi_{i}|$,
$\sum_{i}p_{i}=1$, the elements of the Gram matrix, $\boldsymbol{\mathrm{G}}$,
are defined as 
\begin{equation}
G_{ij}\equiv\sqrt{p_{i}p_{j}}\langle\varphi_{i}|\varphi_{j}\rangle.\label{eq:5}
\end{equation}
We take advantage of the fact that the eigenvalues of $\boldsymbol{\mathrm{G}}$
and $\rho$ are the same (including their multiplicities) \cite{key-27},
and that $\rho^{AE}$ is by construction a mixture of pure states.
In the Appendix A we write explicitly this matrix, which is of dimension
$4\times4$, and compute its exact eigenvalues. Note that the a priori Hilbert space for the joint state 
$\rho^{AE}$ is of dimension 8, given that 4 ancilla states are required
to fully describe the Eve's forward attack, cf. Eq.~\eqref{eq:2}. This entails
a clear advantage of using the Gram matrix over the standard
method (in which $\rho^{AE}$ is written in some orthonormal basis), i.e.: first, finding a suitable orthonormal basis
is not obvious; and second, even if that were the case, computing
the eigenvalues of a $8\times8$ matrix is a considerably more difficult task than
doing it for $4\times4$ matrix. 

From Eq.~\eqref{eq:2}, we deduce that the perturbation that Eve's unitary, $U$, causes
on the qubit state $|0\rangle_{\hat{k}}$ ($|1\rangle_{\hat{k}}$) is given
by $\langle\epsilon_{01}^{\hat{k}}|\epsilon_{01}^{\hat{k}}\rangle$
($\langle\epsilon_{10}^{\hat{k}}|\epsilon_{10}^{\hat{k}}\rangle$).
Considering the generic form of a TWQKD protocol (described in
Section \ref{section:TWQKD} and Fig.~\ref{fig:1}) the set of Bob's prepared states $\{|\psi_{i}\rangle\}_{i}$ can be written as $\{|\psi_{i}\rangle\}_{i}=\{|0\rangle_{\hat{k}},|1\rangle_{\hat{k}}\}_{\hat{k}\in\{\hat{k}\}}$,
where $\{\hat{k}\}$ is the set of Bloch sphere directions used in the Bob's preparation. The fact that Eve has access to the state $\mathbb{I}^{A}/2$
in the forward channel means that Bob prepared $|0\rangle_{\hat{k}}$
and $|1\rangle_{\hat{k}}$ with the same probability, regardless of the preparation basis direction choice, 
$\hat{k}$. As aforementioned, we are assuming that the qubits are
sent through a depolarizing channel and consequently
\begin{equation}
\langle\epsilon_{01}^{\hat{k}}|\epsilon_{01}^{\hat{k}}\rangle=\langle\epsilon_{10}^{\hat{k}}|\epsilon_{10}^{\hat{k}}\rangle=Q_{f},\label{eq:6}
\end{equation}
for any choice $\hat{k}\in\{\hat{k}\}$. Here $Q_{f}$ is the natural noise
of the forward channel, which is simply the probability that Alice
receives $|0\rangle_{\hat{k}}$ ($|1\rangle_{\hat{k}}$) if Bob prepared
$|1\rangle_{\hat{k}}$ ($|0\rangle_{\hat{k}}$). In this way, Eq.~\eqref{eq:6} 
tells us that, in order to pass unnoticed, Eve is restricted
to interactions whose perturbation equals $Q_{f}$ for all the transmitted
states. Considering this constraint, we obtain in Appendix A the following
expressions for the eigenvalues of $\rho^{AE}$: 
\begin{align}
\lambda_{\pm} & =\frac{1}{4}\left[1\pm\sqrt{(1-2\langle\epsilon_{01}^{\hat{z}}|\epsilon_{01}^{\hat{z}}\rangle)^{2}+4|\langle\epsilon_{00}^{\hat{z}}|\epsilon_{10}^{\hat{z}}\rangle|^{2}}\right]\nonumber \\
 & =\frac{1}{4}\left[1\pm\sqrt{(1-2Q_{f})^{2}+4|\langle\epsilon_{00}^{\hat{z}}|\epsilon_{10}^{\hat{z}}\rangle|^{2}}\right],\label{eq:7}
\end{align}
each one with multiplicity 2. Equation \eqref{eq:4} is then translated
into $I_{E}=\underset{\{U\}}{\mathrm{max}}S(\rho^{AE})-1=\underset{|\langle\epsilon_{00}^{\hat{z}}|\epsilon_{10}^{\hat{z}}\rangle|}{\mathrm{max}}S(\rho^{AE})-1$,
for fixed $Q_{f}$. A simple argument (see Appendix
A) leads to conclude that the von Neumann entropy, 
\[
S(\rho^{AE})=-2[\lambda_{+}\mathrm{log}_{2}(\lambda_{+})+\lambda_{-}\mathrm{log}_{2}(\lambda_{-})],
\]
takes its maximum when $|\langle\epsilon_{00}^{\hat{z}}|\epsilon_{10}^{\hat{z}}\rangle|=0$.
Thereby, $\underset{|\langle\epsilon_{00}^{\hat{z}}|\epsilon_{10}^{\hat{z}}\rangle|}{\mathrm{max}}S(\rho^{AE})=1+h(Q_{f})$
and 
\begin{equation}
I_{E}=h(\langle\epsilon_{01}^{\hat{z}}|\epsilon_{01}^{\hat{z}}\rangle)=h(Q_{f}).\label{eq:8}
\end{equation}

In this case, Eq.~\eqref{eq:8} allows to determine Eve's
information in terms of the noise introduced along the $\hat{z}$
direction, $\langle\epsilon_{01}^{\hat{z}}|\epsilon_{01}^{\hat{z}}\rangle$.
This implies that Bob must prepare eigenstates of $\sigma_{z}^{A}$
and Alice must measure such an observable in CM. This kind of protocol 
is clearly non-deterministic since the encoding operations, $\{\mathbb{I}^{A},\sigma_{z}^{A}\}$,
have no effect on the states $\{|0\rangle_{\hat{z}},|1\rangle_{\hat{z}}\}$.
In the next subsection we will see that, regardless of the number
of preparation bases used, Eve's information can not be less than
the given by Eq.~\eqref{eq:8}.

\subsection{Generalization of the security proof to an arbitrary number of preparation
directions}

Here we show that, if the previous protocol is implemented in a depolarizing
channel, no preparation strategy can reduce $I_{E}$ below the value
corresponding to Eq.~\eqref{eq:8}. To this aim we
present an explicit eavesdropping attack that simulates a depolarizing
channel with noise $Q_{f}$ and provides Eve $h(Q_{f})$ bits of information.
Besides the condition $left\langle\epsilon_{01}^{\hat{z}}|\epsilon_{01}^{\hat{z}}\rangle=\langle\epsilon_{10}^{\hat{z}}|\epsilon_{10}^{\hat{z}}\rangle=Q_{f}$,
this attack satisfies the following equations: 
\begin{equation}
\langle\epsilon_{00}^{\hat{z}}|\epsilon_{10}^{\hat{z}}\rangle=\langle\epsilon_{01}^{\hat{z}}|\epsilon_{11}^{\hat{z}}\rangle=0,\label{eq:9.1}
\end{equation}
\begin{equation}
\langle\epsilon_{00}^{\hat{z}}|\epsilon_{01}^{\hat{z}}\rangle=\langle\epsilon_{10}^{\hat{z}}|\epsilon_{11}^{\hat{z}}\rangle=0,\label{eq:9.2}
\end{equation}
\begin{equation}
\langle\epsilon_{01}^{\hat{z}}|\epsilon_{10}^{\hat{z}}\rangle=0,\;\langle\epsilon_{00}^{\hat{z}}|\epsilon_{11}^{\hat{z}}\rangle=1-2Q_{f}.\label{eq:9.3}
\end{equation}

A general qubit pure state may be written as $|\Omega\rangle=\mathrm{sen}\left(\frac{\theta}{2}\right)|0\rangle_{\hat{z}}+e^{i\phi}\mathrm{cos}\left(\frac{\theta}{2}\right)|1\rangle_{\hat{z}},$
where $\theta$ and $\phi$ are the azimuthal and polar angles in the Bloch sphere, respectively.
The effect of the forward attack is then 
\[
U|\Omega\rangle|\epsilon\rangle=|\Omega\rangle|\epsilon(\Omega)\rangle+|\Omega_{\perp}\rangle|\epsilon(\Omega_{\perp})\rangle,
\]
where $|\Omega_{\perp}\rangle=\mathrm{cos}\left(\frac{\theta}{2}\right)|0\rangle_{\hat{z}}-e^{i\phi}\mathrm{sen}\left(\frac{\theta}{2}\right)|1\rangle_{\hat{z}}$
is the state orthogonal to $|\Omega\rangle$ and $\{|\epsilon(\Omega)\rangle,|\epsilon(\Omega_{\perp})\rangle\}$
are ancilla states associated to the Eve's attack. The disturbance generated on $|\Omega\rangle$
is $\langle\epsilon(\Omega_{\perp})|\epsilon(\Omega_{\perp})\rangle=1-\langle\epsilon(\Omega)|\epsilon(\Omega)\rangle$, cf. 
Eq.~\eqref{eq:2.1}. Taking into account Eqs.~\eqref{eq:9.1}-\eqref{eq:9.3},
we show in the Appendix B that $\langle\epsilon(\Omega_{\perp})|\epsilon(\Omega_{\perp})\rangle=Q_{f}$. 

Thus, we have proved that the attack described by Eqs.~\eqref{eq:9.1}-\eqref{eq:9.3}
fully simulates a depolarizing channel with characteristic noise $Q_{f}$.
On the other hand, Eq.~\eqref{eq:9.1} implies that $I_{E}=h(\langle\epsilon_{01}^{\hat{z}}|\epsilon_{01}^{\hat{z}}\rangle)=h(Q_{f})$.
Although this result could suggest that it is sufficient for Bob to
prepare only eigenstates of $\sigma_{z}^{A}$, indeed at least one
additional basis must be incorporated into the preparation. As already
mentioned, this is necessary because the encoding operations $\{\mathbb{I}^{A},\sigma_{z}^{A}\}$
do not affect these prepared states (the eigenstates of $\sigma_{z}^{A}$). Furthermore, the transmission of non-orthogonal
states is the key ingredient to guarantee the protection of the information.
Accordingly, the simplest protocol whose security is established by
Eq.~\eqref{eq:8} could be one where the encoding is performed with the set
$\{\mathbb{I}^{A},\sigma_{z}^{A}\}$ and Bob prepares the states $\{|0\rangle_{\hat{k}},|1\rangle_{\hat{k}}\}_{\hat{k}\in\{\hat{x},\hat{z}\}}$.
The results obtained in this section will be employed to prove the
security of the LM05' protocol in what follows.

\section{Security proof for the LM05' protocol}

The LM05' protocol corresponds specifically to version 2 of the ``Qubit
LM05 protocol, implemented with reverse reconciliation'', proposed
in Ref.~\cite{key-5}. In this scheme the encoding operations are
$\{\mathbb{I}^{A},\sigma_{z}^{A},\sigma_{x}^{A},\sigma_{y}^{A}\}$
and Bob's preparation is carried out with the states $\{|0\rangle_{\hat{k}},|1\rangle_{\hat{k}}\}_{\hat{k}\in\{\hat{x},\hat{z}\}}$.
Moreover, Bob measures the received state in the preparation basis,
interpreting an unchanged state as the bit ``0'' and one flipped
as the bit ``1''. This implies that the protocol is deterministic,
since he can always perfectly distinguish the encoded bit with his
measurements. On the other hand, the operations $\{\sigma_{z}^{A},\sigma_{x}^{A}\}$
can cause a bit flip or act similarly as the identity, depending on the preparation
basis, $\hat{k}\in\{\hat{x},\hat{z}\}$. Therefore, at the end of the quantum transmission Bob must
reveal the preparation bases in order to Alice can infer all the
encoded bits. It is the disclosing of such information what the authors
term ``reverse reconciliation'' in \cite{key-5}. Here we will consider
that Bob's choices and the corresponding to Alice in each mode of the protocol (EM and CM) are
performed with equal probability. In CM she measures randomly the
observables $\sigma_{x}^{A}$ and $\sigma_{z}^{A}$ to determine the
forward noise. 

According to the the description of LM05', there is not a unique Holevo
quantity to be maximized for obtaining Eve's information (Eq.~\eqref{eq:3}).
Instead of that, the states from which Eve can extract the encoding
depend on the associated preparation direction. Suppose for instance
that the Bob's chosen preparation direction was $\hat{z}$. Then, $\rho_{z}^{AE|0}=\frac{1}{4}[U(\mathbb{I}^{A}\otimes|\epsilon\rangle\langle\epsilon|)U^{\dagger}+\sigma_{z}^{A}U(\mathbb{I}^{A}\otimes|\epsilon\rangle\langle\epsilon|)U^{\dagger}\sigma_{z}^{A}]$
and $\rho_{z}^{AE|1}=\frac{1}{4}[\sigma_{x}^{A}U(\mathbb{I}^{A}\otimes|\epsilon\rangle\langle\epsilon|)U^{\dagger}\sigma_{x}^{A}+\sigma_{y}^{A}U(\mathbb{I}^{A}\otimes|\epsilon\rangle\langle\epsilon|)U^{\dagger}\sigma_{y}^{A}]$,
given that the operations sub-set $\{\sigma_{x}^{A},\sigma_{y}^{A}\}$ produce
a bit flip while the sub-set $\{\mathbb{I}^{A},\sigma_{z}^{A}\}$ leave the state
unchanged. In this case the state $\rho^{AE}$ turns out to be 
\begin{equation}
\rho^{AE}=\frac{1}{8}\left(U\mathbb{I}^{A}\otimes|\epsilon\rangle\langle\epsilon|U^{\dagger}+\sum_{w=x,y,z}\sigma_{w}^{A}U\mathbb{I}^{A}\otimes|\epsilon\rangle\langle\epsilon|U^{\dagger}\sigma_{w}^{A}\right)\label{eq:11}
\end{equation}
and the corresponding Holevo quantity,
\begin{equation}
\chi_{z}=S(\rho^{AE})-S(\rho_{z}^{AE|0}).\label{eq:12}
\end{equation}
Equation~\eqref{eq:12} follows from the fact that $\rho_{z}^{AE|1}=\sigma_{x}^{A}\rho_{z}^{AE|0}\sigma_{x}^{A}=\sigma_{y}^{A}\rho_{z}^{AE|0}\sigma_{y}^{A}$
and the von Neumann entropy is invariant under unitary transformations
\cite{key-23}. The same kind of reasoning leads to the Holevo quantity
corresponding to the Bob's preparation in the $\hat{x}$ direction: 
\begin{equation}
\chi_{x}=S(\rho^{AE})-S(\rho_{x}^{AE|0}),\label{eq:13}
\end{equation}
where $\rho_{x}^{AE|0}=\frac{1}{4}[U(\mathbb{I}^{A}\otimes|\epsilon\rangle\langle\epsilon|)U^{\dagger}+\sigma_{x}^{A}U(\mathbb{I}^{A}\otimes|\epsilon\rangle\langle\epsilon|)U^{\dagger}\sigma_{x}^{A}]$.
In this way, Eve's information is given by 
\begin{equation}
I_{E}=\underset{\{U\}}{\mathrm{max}}\frac{1}{2}\left( \chi_{z}+\chi_{x}\right)
\leq\frac{1}{2}\left(
\underset{\{U\}}{\mathrm{max}}\,\chi_{z}
+\underset{\{U\}}{\mathrm{max}}\,\chi_{x}
\right),\label{eq:14}
\end{equation}
where the factor $1/2$ stems from the fact that Bob uses each preparation
basis with equal probability. The upper bound is due to the general
relation $\mathrm{max}[f(x)+g(x)]\leq\mathrm{max}f(x)+\mathrm{max}g(x)$,
for $f$ and $g$ arbitrary functions. 

The optimization problem involved in Eq.~\eqref{eq:14} could, in principle,
be a considerable task. This is so basically due to the fact that the Holevo
quantities are subtractions of two terms, each one depending on the general unitary $U$ employed in the eavesdropping attack.
Moreover, the Gram matrix associated to $\rho^{AE}$ would be of dimension
$8\times8$, which prevents a possible analytical advantage associated to the
use of such a representation. To circumvent this hindrance, we introduce a
slight modification to the protocol, thereby reducing its security
proof to that of the protocol considered in the previous section. Such
a modification consists in additional classical information that Alice
publicly reveals at the end of the quantum transmission, such as follows:
\begin{itemize}
\item If the preparation direction was $\hat{x}$ and Alice used $\mathbb{I}^{A}$
or $\sigma_{z}^{A}$, she announces that $\sigma_{x}^{A}$ and $\sigma_{y}^{A}$
were discarded. Similarly, if Alice used $\sigma_{x}^{A}$ or $\sigma_{y}^{A}$,
she announces that $\mathbb{I}^{A}$ and $\sigma_{z}^{A}$ were discarded. 
\item If the preparation direction was $\hat{z}$ and Alice used $\mathbb{I}^{A}$
or $\sigma_{x}^{A}$, she announces that $\sigma_{y}^{A}$ and $\sigma_{z}^{A}$
were discarded. Similarly, if Alice used $\sigma_{y}^{A}$ or $\sigma_{z}^{A}$,
she announces that $\mathbb{I}^{A}$ and $\sigma_{x}^{A}$ were discarded. 
\end{itemize}
Notice that in this scenario Eve is unable to obtain directly any information
about the encoded bit. However, the encoded states that she can access in
the QBC take a simpler form now. Consider the case when Bob's preparation
was along the $\hat{x}$ direction and Alice used $\mathbb{I}^{A}$.
Thus, we have
\[
\varrho_{x}^{AE|0}=U(\mathbb{I}^{A}/2\otimes|\epsilon\rangle\langle\epsilon|)U^{\dagger},
\]
\begin{equation}
\varrho_{x}^{AE|1}=\sigma_{z}^{A}U(\mathbb{I}^{A}/2\otimes|\epsilon\rangle\langle\epsilon|)U^{\dagger}\sigma_{z}^{A},\label{eq:15}
\end{equation}
where we have designed the density operator by $\varrho$ instead of $\rho$ to
emphasize that these states correspond to the modified protocol. 

The corresponding Holevo quantity, $\chi'_{x}=S(\frac{1}{2}(\varrho_{x}^{AE|0}+\varrho_{x}^{AE|1}))-S(\varrho_{x}^{AE|0})$,
takes exactly the same form when Alice performed $\sigma_{x}^{A}$
or $\sigma_{y}^{A}$, due to the invariance of the von Neumann entropy
under unitary transformations. Using the same reasoning, it is easy
to obtain the Holevo quantity corresponding to the case when the preparation
direction is $\hat{z}$, whose the general expression is given by
\begin{equation}
\chi'_{w}=S(\varrho_{w}^{AE})-S(\varrho^{AE|0});\quad w=z,x,\label{eq:16}
\end{equation}
with $\varrho^{AE|0}=\varrho_{x}^{AE|0}$ and $\varrho_{w}^{AE}=\frac{1}{4}[U(\mathbb{I}^{A}\otimes|\epsilon\rangle\langle\epsilon|)U^{\dagger}+\sigma_{w'}^{A}U(\mathbb{I}^{A}\otimes|\epsilon\rangle\langle\epsilon|)U^{\dagger}\sigma_{w'}^{A}]$;
$w'$ being the complementary direction of $w$, i.e., $w'=x$ if $w=z$ and vice-versa.
The associated expression for Eve's information, $\mathcal{I}_{E}$,
is the same of Eq.~\eqref{eq:14} with the modified Holevo quantities.
The bound appearing in that equation can be written as 
\[
\mathcal{I}_{E}\leq\frac{1}{2}(\mathcal{I}_{E}^{z}+\mathcal{I}_{E}^{x}),
\]
where $\mathcal{I}_{E}^{w}=\underset{\{U\}}{\mathrm{max}}\chi'_{w}$.
It is worthwhile to note that $\mathcal{I}_{E}^{x}$ corresponds
exactly to $I_{E}$ in Eq.~\eqref{eq:8} and hence $\mathcal{I}_{E}^{x}=h(\langle\epsilon_{01}^{\hat{z}}|\epsilon_{01}^{\hat{z}}\rangle)$.
To find the expression for $\mathcal{I}_{E}^{z}$, we can proceed
in a way completely analogous to the one leading to Eq.~\eqref{eq:8},
writing the Gram matrix with the states $\{|\epsilon_{ij}^{\hat{x}}\rangle\}$
instead of $\{|\epsilon_{ij}^{\hat{z}}\rangle\}$ (see Eq.~\eqref{eq:2}
and Appendix A). It is thus not difficult to grasp that the result
is $\mathcal{I}_{E}^{z}=h(\langle\epsilon_{01}^{\hat{x}}|\epsilon_{01}^{\hat{x}}\rangle)$.
In this way, the estimation of the noise $\langle\epsilon_{01}^{\hat{z}}|\epsilon_{01}^{\hat{z}}\rangle$
($\langle\epsilon_{01}^{\hat{x}}|\epsilon_{01}^{\hat{x}}\rangle$)
allows to determine Eve's information when the encoded states were
prepared along the $\hat{x}$ ($\hat{z}$) direction. Assuming a depolarizing
channel we get 
\begin{equation}
\mathcal{I}_{E}\leq\frac{1}{2}[h(\langle\epsilon_{01}^{\hat{x}}|\epsilon_{01}^{\hat{x}}\rangle)+h(\langle\epsilon_{01}^{\hat{z}}|\epsilon_{01}^{\hat{z}}\rangle)]=h(Q_{f}).\label{eq:17}
\end{equation}

At this point, we remark that the objective of the introduced modification
is to simplify the security proof and by no means is necessary in
practice. Actually, it potentially reduces Eve's uncertainty about
the encoding and therefore $\mathcal{I}_{E}$ (Eq.~\eqref{eq:17}) is an upper bound to
the Eve's information, $I_{E}$, for the LM05' protocol. Furthermore, such
a bound coincides with the one calculated in \cite{key-5} employing other methods. On the
other hand, we show in Appendix~C that the attack represented by Eqs.~\eqref{eq:9.1}-\eqref{eq:9.3} makes such bound in Eq.~\eqref{eq:17} tight,
which implies that for the modified version Eve can in fact get $h(Q_{f})$
bits of information.

\section{Other deterministic TWQKD protocols}

\subsection{The TWQKD six-state protocol}

\begin{figure}
\includegraphics[scale=0.04015]{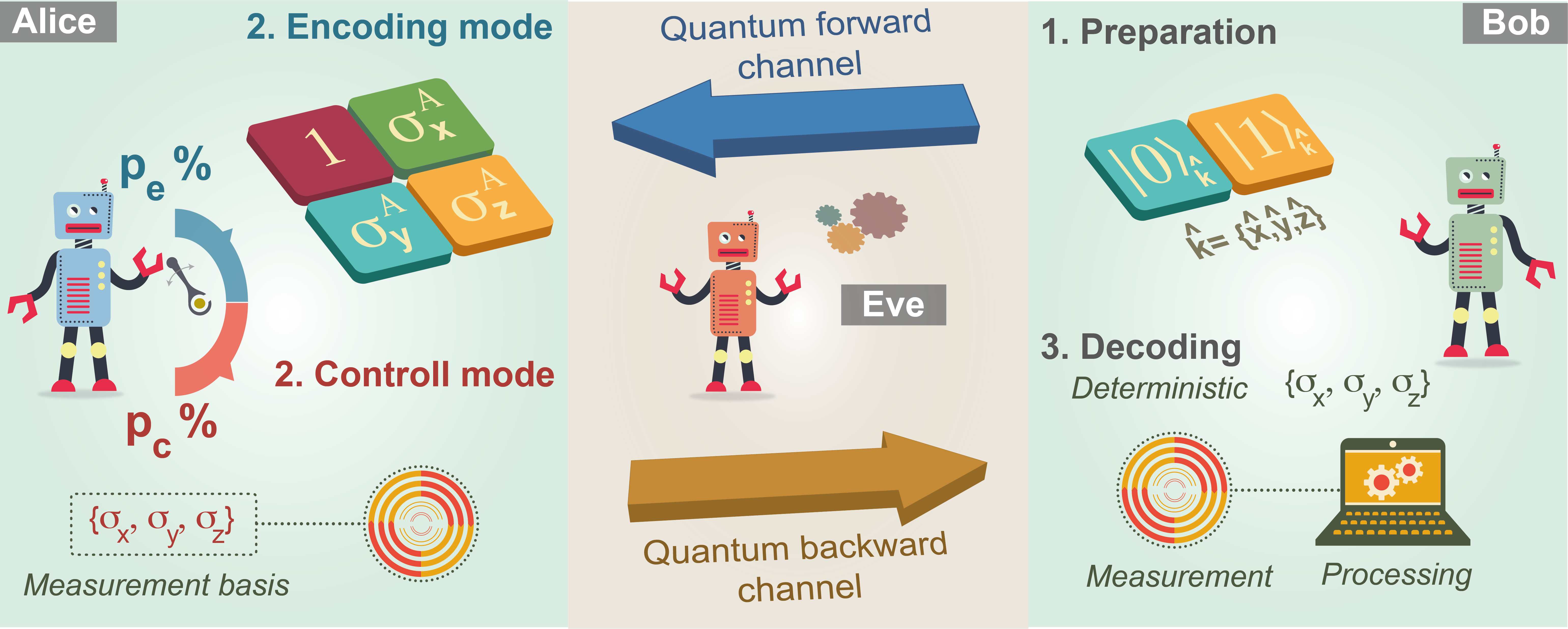}
\caption{(Color online) Schematic illustration of the quantum part of the TWQKD six-state protocol (the classical communication channel is omitted). Alice (on the left side) and Bob (on the right side) perform each of their possible choices with equal probability, 
i.e., basis direction $\hat{k}$ for the state preparation, (in EM) Alice's encoding operation, and (in CM) Alice's measurement basis.}
\label{fig:2}
\end{figure}

Considering the LM05' protocol, we note that if besides $\hat{z}$
and $\hat{x}$ one includes the preparation direction $\hat{y}$ and
Bob always measures in the preparation basis, the resulting scheme
is still deterministic. The reason is that, as in the case of LM05',
the encoding operations $\{\mathbb{I}^{A},\sigma_{z}^{A},\sigma_{x}^{A},\sigma_{y}^{A}\}$
always flip or leave unchanged the prepared state depending on the preparation basis. Now let us suppose
that Alice and Bob perform their choices according to a uniform probability
distribution. The resulting scheme, which we term ``TWQKD six-state''
(given the use of the same preparation strategy of the six-state protocol
\cite{key-2}), had not been proposed to date, to the best of our knowledge. We
leave the remaining details of the protocol identical to those of
LM05'. Regarding the decoding, for example, Alice and Bob agree that
a unchanged (flipped) state corresponds to the bit ``0'' (``1'').
To that aim Bob must publicly disclose his basis preparation choices after the
quantum transmission, so that Alice can know the effect of her operations
in each case. Furthermore, in CM Alice should measure any of the observables
$\{\sigma_{x}^{A},\sigma_{y}^{A},\sigma_{z}^{A}\}$ with equal probability,
in order to determine the noise introduced along each direction. Figure~\ref{fig:2} illustrates the quantum part of this protocol.

Similar to the LM05' protocol case, for the TWQKD six-state protocol
there is a Holevo quantity associated to each possible direction chosen
by Bob:
\begin{equation}
\chi_{w}=S(\rho^{AE})-S(\rho_{w}^{AE|0});\quad w=x,y,z,\label{eq:32}
\end{equation}
where the Alice-Eve joint state $\rho^{AE}$ is given by Eq.~\eqref{eq:11} with the conditional joint state $\rho_{w}^{AE|0}=\frac{1}{4}[U(\mathbb{I}^{A}\otimes|\epsilon\rangle\langle\epsilon|)U^{\dagger}+\sigma_{w}^{A}U(\mathbb{I}^{A}\otimes|\epsilon\rangle\langle\epsilon|)U^{\dagger}\sigma_{w}^{A}]$, $w=x,y,z$.
Likewise, the optimization to obtain Eve's information, $I_{E}=\underset{\{U\}}{\mathrm{max}}\frac{1}{3}\sum_{w=x,y,z}\chi_{w}$,
is at least as difficult as doing it for LM05' (Eq.~\eqref{eq:14}).
Nevertheless, the security of the TWQKD six-state scheme immediately
follows from that of LM05', since the only difference is the usage
of the additional preparation direction $\hat{y}$ and it constrains
more Eve's strategies (she must now cause the same perturbation not
only along $\hat{x}$ and $\hat{z}$ but also along $\hat{y}$). This
implies that Eve's information is also upper bounded by $h(Q_{f})$,
according to Eq.~\eqref{eq:17}. 

\begin{figure}
\includegraphics[scale=0.60]{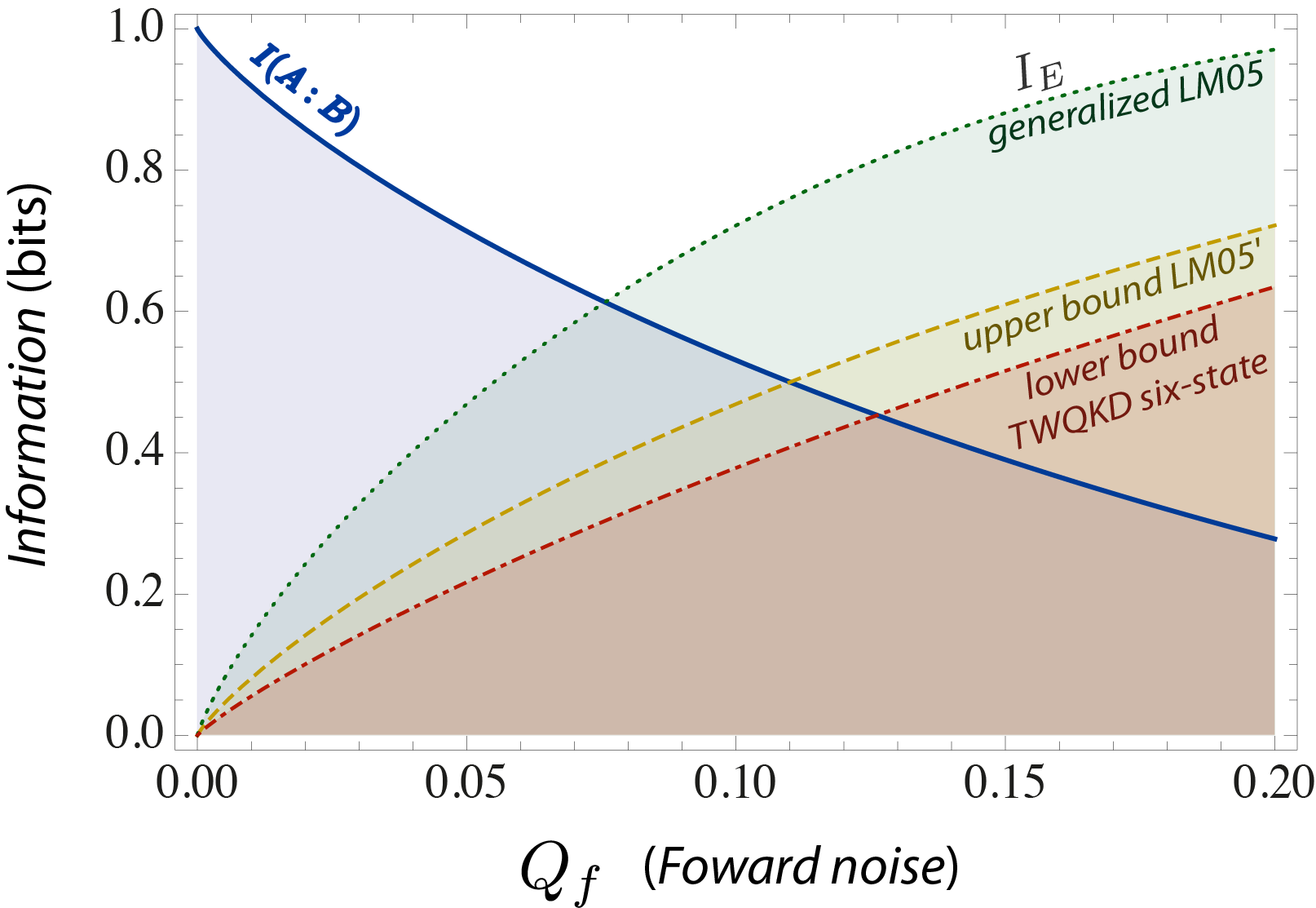}
\caption{(Color online) Comparison of the Alice-Bob Mutual Information $I(A:B)$ (blue solid line) and Eve's Information ($I_E$)  for different protocols, as function of the forward noise ($Q_f$). It is assumed that the overall noise (Q) and the forward noise coincide, which is
the case if the QFC and the QBC are correlated (see, e.g., in Ref.
\cite{key-5} some comments at this respect).$I_E$ for the generalized LM05 protocol is plotted as a (dark-green) dotted line. The upper bound to $I_E$ for LM05' protocol is represented by a (dark-yellow) dashed line. The lower bound to $I_E$ for the TWQKD six-state is drawn as the (dark-red) dash-dotted line.}
\label{fig:5.2}
\end{figure}

In Appendix C we compute the von Neumann entropies, $S(\rho^{AE})$
and $S(\rho_{w}^{AE|0})$, for the unitary transformation described by Eqs.
\eqref{eq:9.1}-\eqref{eq:9.3}. The corresponding eavesdropping information
is given by 
\begin{equation}
\frac{1}{3}\sum_{w=x,y,z}\chi_{w}=Q_{f}+(1-Q_{f})h\left(\frac{2-3Q_{f}}{2(1-Q_{f})}\right).\label{eq:33}
\end{equation}
The right hand side of Eq.~\eqref{eq:33} is identical to the expression for Eve's
information corresponding to the six-state protocol \cite{scarani09,key-15,key-17}.
This is remarkable if one considers that the same kind of states are
prepared in both the six-state scheme and its TWQKD version. Moreover,
the analysed attack is in fact one of the most powerful since it makes
tight the bound in Eq.~\eqref{eq:17}, as we also prove in Appendix C. Based
on these facts, we conjecture that Eq.~\eqref{eq:33} actually corresponds
to $I_{E}$ for the TWQKD six-state protocol. We can summarize the
results concerning the LM05' and TWQKD six-state schemes with the
following inequallity: 
\begin{align}
Q_{f}+(1-Q_{f})h\left(\frac{2-3Q_{f}}{2(1-Q_{f})}\right) &\leq I_{E}^{\textrm{TWQKD six-state}} \nonumber \\
&\leq I_{E}^{\textrm{LM05'}}\leq h(Q_{f}).\label{eq:34}
\end{align}
These bounds are plotted in Fig.~\ref{fig:5.2}, as well as the mutual information between Alice and Bob ($I(A:B)$), and the eavesdropping information ($I_E$) for the
LM05 protocol.

\subsection{Generalized version of the LM05 protocol }

Now let us consider a TWQKD protocol in which Bob prepares states from
the set $\{|\psi_{i}\rangle\}_{i}=\left\{ \frac{1}{\sqrt{2}}\left(|0\rangle+e^{i\phi}|1\rangle\right)\right\} _{0\leq\phi\leq2\pi}$,
i.e., any state on the perpendicular plane to the $z$ direction
of the Bloch sphere, and Alice performs the encoding with $\{U_{i}^{A}\}_{i}=\{\mathbb{I}^{A},\sigma_{z}^{A}\}$.
Such a protocol is clearly deterministic and can be considered as
a generalization of the LM05 protocol \cite{key-4}, which only uses
two bases for the preparation, e.g. $\sigma_{x}$ and $\sigma_{y}$.
We will show that this generalized protocol does not have any advantage over the standard
LM05, within the assumption that the channel used to transmit the
states is depolarizing. We start computing the eavesdropping
information for the attack described by the following conditions:
\begin{equation}
\langle\epsilon_{00}^{\hat{z}}|\epsilon_{10}^{\hat{z}}\rangle=\langle\epsilon_{01}^{\hat{z}}|\epsilon_{11}^{\hat{z}}\rangle=0,\label{eq:9.1-1}
\end{equation}
\begin{equation}
\langle\epsilon_{00}^{\hat{z}}|\epsilon_{01}^{\hat{z}}\rangle=\langle\epsilon_{10}^{\hat{z}}|\epsilon_{11}^{\hat{z}}\rangle=0,\label{eq:9.2-1}
\end{equation}
\begin{equation}
\langle\epsilon_{01}^{\hat{z}}|\epsilon_{10}^{\hat{z}}\rangle=0,\;\langle\epsilon_{00}^{\hat{z}}|\epsilon_{11}^{\hat{z}}\rangle=1-\langle\epsilon_{01}^{\hat{z}}|\epsilon_{01}^{\hat{z}}\rangle.\label{eq:9.3-1}
\end{equation}
This attack only differs from the one represented by Eqs.
\eqref{eq:9.1}-\eqref{eq:9.3} concerning the value of the element $\langle\epsilon_{00}^{\hat{z}}|\epsilon_{11}^{\hat{z}}\rangle$.
In addition, we write $\langle\epsilon_{01}^{\hat{z}}|\epsilon_{01}^{\hat{z}}\rangle$
instead of $Q_{f}$ because $\langle\epsilon_{01}^{\hat{z}}|\epsilon_{01}^{\hat{z}}\rangle$
is not the disturbance measured in this protocol. 

Considering that the state $|\phi\rangle=\frac{1}{\sqrt{2}}\left(|0\rangle+e^{i\phi}|1\rangle\right)$
is equal to $|\Omega\rangle$ (defined in Section III B) for $\theta=\pi/2$,
we can use Eq.~\eqref{eq:B.3} from Appendix B to compute the perturbation
that the attack defined by Eqs.~\eqref{eq:9.1-1}-\eqref{eq:9.3-1} introduces
on such a state. Expressing the attack unitary operation as $U|\phi\rangle|\epsilon\rangle=|\phi\rangle|\epsilon(\phi)\rangle+|\phi_{\perp}\rangle|\epsilon(\phi_{\perp})\rangle$,
with $|\phi_{\perp}\rangle$ being the state perpendicular to $|\phi\rangle$,
we have
\begin{align*}
\langle\epsilon(\phi)|\epsilon(\phi)\rangle & =\mathrm{sen}^{4}\left(\frac{\pi}{4}\right)(1-\langle\epsilon_{01}^{\hat{z}}|\epsilon_{01}^{\hat{z}}\rangle)\\
&+\mathrm{sen}^{2}\left(\frac{\pi}{4}\right)\mathrm{cos}^{2}\left(\frac{\pi}{4}\right)\langle\epsilon_{00}^{\hat{z}}|\epsilon_{11}^{\hat{z}}\rangle\\
&+2\mathrm{sen}^{2}\left(\frac{\pi}{4}\right)\mathrm{cos}^{2}\left(\frac{\pi}{4}\right)\langle\epsilon_{01}^{\hat{z}}|\epsilon_{01}^{\hat{z}}\rangle\\
&+\mathrm{cos}^{2}\left(\frac{\pi}{4}\right)\mathrm{sen}^{2}\left(\frac{\pi}{4}\right)\langle\epsilon_{11}^{\hat{z}}|\epsilon_{00}^{\hat{z}}\rangle\\
&+\mathrm{cos}^{4}\left(\frac{\pi}{4}\right)(1-\langle\epsilon_{01}^{\hat{z}}|\epsilon_{01}^{\hat{z}}\rangle)=1-\frac{\langle\epsilon_{01}^{\hat{z}}|\epsilon_{01}^{\hat{z}}\rangle}{2},
\end{align*}
where Eqs.~\eqref{eq:9.1-1}-\eqref{eq:9.3-1} have been employed. Thus,
using the unitarity condition, we have $\langle\epsilon(\phi_{\bot})|\epsilon(\phi_{\bot})\rangle=\langle\epsilon_{01}^{\hat{z}}|\epsilon_{01}^{\hat{z}}\rangle/2$.

Using this result and the fact that, from Eq.~\eqref{eq:9.1-1}, this
attack provides Eve $h(\langle\epsilon_{01}^{\hat{z}}|\epsilon_{01}^{\hat{z}}\rangle)$
bits of information (it fulfills the condition to get Eq.~\eqref{eq:8}),
we get
\begin{equation}
I_{E}=h\left(2\langle\epsilon(\phi_{\bot})|\epsilon(\phi_{\bot})\rangle\right)=h(2Q_{f}).\label{eq:C.4}
\end{equation}
We have taken into account that $\langle\epsilon(\phi_{\bot})|\epsilon(\phi_{\bot})\rangle$
must coincide with the forward noise. The expression in Eq.~\eqref{eq:C.4}
is exactly the found for the LM05 protocol in \cite{key-10} and therefore
we have proved that it is not possible to improve the protocol security with
this extended version on the number of preparation basis.

\section*{Conclusions}

We have studied the performance of TWQKD protocols that distribute
a secret key by means of non-orthogonal qubit states. For the security
analysis we have used techniques whose application is new in the
context of secure communications, to the best of our knowledge. Specifically, we employed a
matrix representation \cite{key-27} that allows to easily compute 
the eigenvalues of certain density operators, involved in the determination
of the amount of the eavesdropping information. In this way, a simple calculation
leads to Eq.~\eqref{eq:8}, which is the more important expression for
the posterior analysis. With this equation we provided a new (alternative) security
proof for a protocol proposed in Ref. \cite{key-5} (that we have
named LM05'), obtaining an upper bound to the eavesdropping information
that coincides with the previous result in \cite{key-5}. Moreover, we proposed
a novel protocol (TWQKD six-state) that is at least as secure as LM05' and
computed an analytical lower bound to the eavesdropping information. 
Such a bound equals the maximum eavesdropping information corresponding to the six-state scheme \cite{scarani09,key-15,key-17}. 
We remark that the security of the TWQKD six-state protocol is possibly better than that of the LM05' protocol, since an additional
preparation basis is used by Bob, and therefore our lower bound could be tight.   
Finally, we showed that the inclusion of more than two preparation
bases does not improve the security of the LM05 protocol \cite{key-10}.
The construction of a full security proof for the new TWQKD six-state
protocol proposed here as well as its experimental analysis and tests are interesting directions for a future research. 

\noindent

{\bf Acknowledgments.}
We thank T. B. Batalh\~{a}o, S. P. Walborn, and L. S. Cruz for valuable discussions and suggestions. We acknowledge financial support from UFABC, CNPq, CAPES, and FAPESP. RMS acknowledges support from the Royal Society through the Newton Advanced Fellowship scheme (grant n.~R1660101).  This work was performed as part of the Brazilian National Institute of Science and Technology for Quantum Information (INCT-IQ). 

\appendix

\section{von Neumann entropy of the state $\rho^{AE}$ (Eq.~\eqref{eq:4})}

In this appendix we maximize the von Neumann entropy for the Alice-Eve joint state
$\rho^{AE}=\frac{1}{4}\left[ U(\mathbb{I}\otimes|\epsilon\rangle\langle\epsilon|)U^{\dagger}+\sigma_{z}^{A}U(\mathbb{I}\otimes|\epsilon\rangle\langle\epsilon|)U^{\dagger}\sigma_{z}^{A}\right]$,
assuming that $\langle\epsilon_{01}^{\hat{z}}|\epsilon_{01}^{\hat{z}}\rangle=\langle\epsilon_{10}^{\hat{z}}|\epsilon_{10}^{\hat{z}}\rangle=Q_{f}$
is fixed. Using Eq.~\eqref{eq:2}, with $\hat{k}=\hat{z}$, and 
Eq.~\eqref{eq:5}, the corresponding Gram matrix is explicitly written as
\[
\boldsymbol{\mathrm{G}}=\frac{1}{4}\begin{pmatrix}1 & 0 & 1-2Q_{f} & 2\langle\epsilon_{00}^{\hat{z}}|\epsilon_{10}^{\hat{z}}\rangle\\
0 & 1 & 2\langle\epsilon_{10}^{\hat{z}}|\epsilon_{00}^{\hat{z}}\rangle & 2Q_{f}-1\\
1-2Q_{f} & 2\langle\epsilon_{00}^{\hat{z}}|\epsilon_{10}^{\hat{z}}\rangle & 1 & 0\\
2\langle\epsilon_{10}^{\hat{z}}|\epsilon_{00}^{\hat{z}}\rangle & 2Q_{f}-1 & 0 & 1
\end{pmatrix}.
\]
The eigenvalues of $\boldsymbol{\mathrm{G}}$ are given by
\[
\lambda_{\pm}=\frac{1}{4}[1\pm\sqrt{(1-2Q_{f})^{2}+4|\langle\epsilon_{00}^{\hat{z}}|\epsilon_{10}^{\hat{z}}\rangle|^{2}}],
\]
which implies that the von Neumann entropy $S[\rho^{AE}]$ will be
a function of $Q_{f}$ and $|\langle\epsilon_{00}^{\hat{z}}|\epsilon_{10}^{\hat{z}}\rangle|$.
For a fixed $Q_{f}$ we obtain 
\[
S(\rho^{AE})=-2[\lambda_{+}(x)\mathrm{log}_{2}(\lambda_{+}(x))+\lambda_{-}(x)\mathrm{log}_{2}(\lambda_{-}(x))],
\]
with $x\equiv|\langle\epsilon_{00}^{\hat{z}}|\epsilon_{10}^{\hat{z}}\rangle|$.
Taking the derivative with respect $x$, it is easy to see that
$S(\rho^{AE})$ is monotonically decreasing in the interval $(0,1)$
and therefore $S(\rho^{AE})$ takes its maximum at $x=0$.

\section{Disturbance on any qubit pure state caused by the attack described by Eqs.~\eqref{eq:9.1}-\eqref{eq:9.3}}

Using the expression for a general qubit pure state, $|\Omega\rangle=\mathrm{sen}\left(\frac{\theta}{2}\right)|0\rangle_{\hat{z}}+e^{i\phi}\mathrm{cos}\left(\frac{\theta}{2}\right)|1\rangle_{\hat{z}}$,
given in Section III B, we can write the action of an eavesdropping
interaction $U$ as 
\begin{align}
U(|\Omega\rangle|\epsilon\rangle) & =\mathrm{sen}\left(\frac{\theta}{2}\right)U|0\rangle_{\hat{z}}|\epsilon\rangle+e^{i\phi}\mathrm{cos}\left(\frac{\theta}{2}\right)U|1\rangle_{\hat{z}}|\epsilon\rangle \nonumber \\
 & =\mathrm{sen}\left(\frac{\theta}{2}\right)\left\{ |0\rangle_{\hat{z}}|\epsilon_{00}^{\hat{z}}\rangle+|1\rangle_{\hat{z}}|\epsilon_{01}^{\hat{z}}\rangle\right\} \nonumber \\
 &+e^{i\phi}\mathrm{cos}\left(\frac{\theta}{2}\right)\left\{ |0\rangle_{\hat{z}}|\epsilon_{10}^{\hat{z}}\rangle+|1\rangle_{\hat{z}}|\epsilon_{11}^{\hat{z}}\rangle\right\} ,\label{eq:B.1}
\end{align}
were we have considered the linearity of $U$ and Eq.~\eqref{eq:2}
has been used, with $\hat{k}=\hat{z}$. 

Taking into account that $|\Omega\rangle$ and its orthonormal state,
$|\Omega_{\bot}\rangle=\mathrm{cos}\left(\frac{\theta}{2}\right)|0\rangle_{\hat{z}}-e^{i\phi}\mathrm{sen}\left(\frac{\theta}{2}\right)|1\rangle_{\hat{z}}$,
are defined in terms of the states $\{|0\rangle_{\hat{z}},|1\rangle_{\hat{z}}\}$,
we have
\[
|0\rangle_{\hat{z}}=\mathrm{sen}\left(\frac{\theta}{2}\right)|\Omega\rangle+\mathrm{cos}\left(\frac{\theta}{2}\right)|\Omega_{\bot}\rangle,
\]
\begin{equation}
|1\rangle_{\hat{z}}=e^{-i\phi}\left(\mathrm{cos}\left(\frac{\theta}{2}\right)|\Omega\rangle-\mathrm{sen}\left(\frac{\theta}{2}\right)|\Omega_{\bot}\rangle\right).\label{eq:B.2}
\end{equation}
Therefore, using Eq.~\eqref{eq:B.2}, we can rewrite Eq.~\eqref{eq:B.1} as
\[
U|\Omega\rangle|\epsilon\rangle=|\Omega\rangle|\epsilon(\Omega)\rangle+|\Omega_{\perp}\rangle|\epsilon(\Omega_{\perp})\rangle,
\]
with 
\begin{align}
|\epsilon(\Omega)\rangle & =\mathrm{sen}^{2}\left(\frac{\theta}{2}\right)|\epsilon_{00}^{\hat{z}}\rangle+\mathrm{sen}\left(\frac{\theta}{2}\right)\mathrm{cos}\left(\frac{\theta}{2}\right)e^{-i\phi}|\epsilon_{01}^{\hat{z}}\rangle \nonumber \\
& +e^{i\phi}\mathrm{cos}\left(\frac{\theta}{2}\right)\mathrm{sen}\left(\frac{\theta}{2}\right)|\epsilon_{10}^{\hat{z}}\rangle+\mathrm{cos}^{2}\left(\frac{\theta}{2}\right)|\epsilon_{11}^{\hat{z}}\rangle,\label{eq:B.3}
\end{align}
and a similar expression holds for $|\epsilon(\Omega_{\bot})\rangle$. 

If the states $\{|\epsilon_{ij}^{\hat{z}}\rangle\}$ satisfy Eqs.
\eqref{eq:9.1}-\eqref{eq:9.3}, then
\begin{align*}
\langle\epsilon(\Omega)|\epsilon(\Omega)\rangle & =\mathrm{sen}^{4}\left(\frac{\theta}{2}\right)(1-Q_{f})+\mathrm{sen}^{2}\left(\frac{\theta}{2}\right)\mathrm{cos}^{2}\left(\frac{\theta}{2}\right)\langle\epsilon_{00}^{\hat{z}}|\epsilon_{11}^{\hat{z}}\rangle \nonumber\\
& +2\mathrm{sen}^{2}\left(\frac{\theta}{2}\right)\mathrm{cos}^{2}\left(\frac{\theta}{2}\right)Q_{f}\nonumber \\
 & +\mathrm{cos}^{2}\left(\frac{\theta}{2}\right)\mathrm{sen}^{2}\left(\frac{\theta}{2}\right)\langle\epsilon_{11}^{\hat{z}}|\epsilon_{00}^{\hat{z}}\rangle \nonumber \\
 &+\mathrm{cos}^{4}\left(\frac{\theta}{2}\right)(1-Q_{f})=1-Q_{f}.
\end{align*}
Hence, by the unitarity of $U$, $\langle\epsilon(\Omega_{\bot})|\epsilon(\Omega_{\bot})\rangle=1-\langle\epsilon(\Omega)|\epsilon(\Omega)\rangle=Q_{f}.$

\section{Probable expression for Eve's information in the TWQKD six-state
protocol}

Here we compute the eavesdropping information for the TWQKD six-state
protocol, when the attack is given by Eqs.~\eqref{eq:9.1}-\eqref{eq:9.3}.
The average over the associated Holevo quantities, Eq.~\eqref{eq:32},
is:
\[
\frac{1}{3}\sum_{w=x,y,z}\chi_{w}=\frac{1}{3}\sum_{w=x,y,z}\left\{ S\left[\rho^{AE}\right]-S\left[\rho_{w}^{AE|0}\right]\right\} .
\]

The elements of the Gram Matrix corresponding to $\rho^{AE}$ result
from the scalar products between the states $\{U|i\rangle_{\hat{z}}|\epsilon\rangle,\sigma_{w}^{A}U|i\rangle_{\hat{z}}|\epsilon\rangle\}_{i,w}$,
being $U$ the attack described by Eqs.~\eqref{eq:9.1}-\eqref{eq:9.3}.
For this attack it turns out that the sets $\{|\varphi_{1}^{(1)}\rangle,...,|\varphi_{4}^{(1)}\rangle\}\equiv\{U|0\rangle_{\hat{z}}|\epsilon\rangle,\sigma_{x}^{A}U|1\rangle_{\hat{z}}|\epsilon\rangle,\sigma_{y}^{A}U|1\rangle_{\hat{z}}|\epsilon\rangle,\sigma_{z}^{A}U|0\rangle_{\hat{z}}|\epsilon\rangle\}$
and $\{|\varphi_{1}^{(2)}\rangle,...,|\varphi_{4}^{(2)}\rangle\}\equiv\{U|1\rangle_{\hat{z}}|\epsilon\rangle,\sigma_{x}^{A}U|0\rangle_{\hat{z}}|\epsilon\rangle,\sigma_{y}^{A}U|0\rangle_{\hat{z}}|\epsilon\rangle,\sigma_{z}^{A}U|1\rangle_{\hat{z}}|\epsilon\rangle\}$
are orthogonal. We denote the Gram Matrices associated to the first
and second sets as $\mathbf{G^{(1)}}$and $\mathbf{G^{(2)}}$, respectively,
and their elements are (see Eq.~\eqref{eq:5}) $G_{ij}^{(k)}=\frac{1}{4}\langle\varphi_{i}^{(k)}|\varphi_{j}^{(k)}\rangle$,
$k=1,2$. The orthogonality property implies that the eigenvalues
of $\rho^{AE}$ are derived from the corresponding to $\mathbf{G^{(1)}}$and
$\mathbf{G^{(2)}}$. Using Eqs.~\eqref{eq:9.1}-\eqref{eq:9.3} and
the usual assumption of the depolarizing channel, we get 
\[
\mathbf{G^{(1)}}=\frac{1}{4}\begin{pmatrix}1 & 1-2Q_{f} & -i(1-2Q_{f}) & 1-2Q_{f}\\
1-2Q_{f} & 1 & -i(1-2Q_{f}) & 1-2Q_{f}\\
i(1-2Q_{f}) & i(1-2Q_{f}) & 1 & i(1-2Q_{f})\\
1-2Q_{f} & 1-2Q_{f} & -i(1-2Q_{f}) & 1
\end{pmatrix},
\]
\[
\mathbf{G^{(2)}}=\frac{1}{4}\begin{pmatrix}1 & 1-2Q_{f} & i(1-2Q_{f}) & -(1-2Q_{f})\\
1-2Q_{f} & 1 & i(1-2Q_{f}) & -(1-2Q_{f})\\
-i(1-2Q_{f}) & -i(1-2Q_{f}) & 1 & i(1-2Q_{f})\\
-(1-2Q_{f}) & -(1-2Q_{f}) & -i(1-2Q_{f}) & 1
\end{pmatrix}.
\]
One finds that the eigenvalues
of $\mathbf{G^{(1)}}$ and $\mathbf{G^{(2)}}$ are $\{1-1.5Q_{f},0.5Q_{f}\}$,
the first one having multiplicity 3. Therefore, 
\begin{equation}
S(\rho^{AE})=2-\frac{3}{2}Q_{f}\mathrm{log}_{2}Q_{f}-\frac{2-3Q_{f}}{2}\mathrm{log}_{2}(2-3Q_{f}).\label{eq:D.1}
\end{equation}

For the analysed attack we already know that $S\left(\rho_{z}^{AE|0}\right)=1+h(\langle\epsilon_{01}^{\hat{z}}|\epsilon_{01}^{\hat{z}}\rangle)=1+h(Q_{f})$,
according to Eq.~\eqref{eq:9.1}. On the other hand, the sets $\{U|i\rangle_{\hat{z}}|\epsilon\rangle,\sigma_{x}^{A}U|i\rangle_{\hat{z}}|\epsilon\rangle\}_{i}$
and $\{U|i\rangle_{\hat{z}}|\epsilon\rangle,\sigma_{y}^{A}U|i\rangle_{\hat{z}}|\epsilon\rangle\}_{i}$
determine the Gram matrices needed to compute $S\left[\rho_{x}^{AE|0}\right]$
and $S\left[\rho_{y}^{AE|0}\right]$. The obtained expressions are
\[
\mathbf{G_{x}}=\frac{1}{4}\begin{pmatrix}1 & 0 & 0 & 1-2Q_{f}\\
0 & 1 & 1-2Q_{f} & 0\\
0 & 1-2Q_{f} & 1 & 0\\
1-2Q_{f} & 0 & 0 & 1
\end{pmatrix},
\]
corresponding to $\rho_{x}^{AE|0}$, and 
\[
\mathbf{G_{y}}=\frac{i}{4}\begin{pmatrix}-i & 0 & 0 & -(1-2Q_{f})\\
0 & -i & (1-2Q_{f}) & 0\\
0 & -(1-2Q_{f}) & -i & 0\\
(1-2Q_{f}) & 0 & 0 & -i
\end{pmatrix},
\]
for $\rho_{y}^{AE|0}$. The eigenvalues of $\mathbf{G_{x}}$ and $\mathbf{G_{y}}$
are $\{0.5Q_{f},0.5(1-Q_{f})\}$, each one with multiplicity 2. Therefore,
\begin{equation}
S\left(\rho_{w}^{AE|0}\right)=1+h(Q_{f});\quad w=x,y,z.\label{eq:D.2}
\end{equation}

From Eqs.~\eqref{eq:D.1} and \eqref{eq:D.2}, the information that
Eve gets using this attack is thus 
\[
\frac{1}{3}\sum_{w=x,y,z}\chi_{w}=Q_{f}+(1-Q_{f})h\left(\frac{2-3Q_{f}}{2(1-Q_{f})}\right).
\]
Notice in particular that Eq.~\eqref{eq:D.2} entails that the bound
Eq.~\eqref{eq:17} is tight.

\end{document}